\documentstyle[11pt,epsf]{article}
 1
 1
 1

\def\lsim{\mathrel{\raise.3ex\hbox{$<$\kern-.75em\lower1ex\hbox{$\sim$}}}}
\def\gsim{\mathrel{\raise.3ex\hbox{$>$\kern-.75em\lower1ex\hbox{$\sim$}}}}
\begin{document}
\noindent
\thispagestyle{empty}
\renewcommand{\thefootnote}{\fnsymbol{footnote}}
\begin{flushright}
{\bf UCSD/PTH 97-08}\\
{\bf TTP96-45}\\
{\bf hep-ph/9703404}\\
{\bf March 1997}\\
\end{flushright}
\vspace{.5cm}
\begin{center}
  \begin{Large}\bf
Two-Loop Corrections to the Electromagnetic Vertex 
\\[1mm]
for Energies close to Threshold
  \end{Large}
  \vspace{1.5cm}

\begin{large}
 A.H. Hoang
\end{large}
\begin{center}
\begin{it}
   Department of Physics,
   University of California, San Diego,\\
   La Jolla, CA 92093--0319, USA\\ 
\end{it} 
\end{center}

  \vspace{4cm}
  {\bf Abstract}\\
\vspace{0.3cm}

\noindent
\begin{minipage}{14.5cm}
\begin{small}
Two-loop contributions to the electromagnetic form factors are
calculated in the kinematic regime close to the fermion-antifermion
threshold. The results are presented in an expansion in the 
velocity $\beta$ of the fermions in the c.m.~frame up to
next-to-next-to
leading order in $\beta$. The existence of a new Coulomb singularity 
logarithmic in $\beta$, which is closely related to the 
${\cal{O}}(\alpha^2\ln\alpha)$ corrections known from positronium
decays, is demonstrated. It is shown that due to this Coulomb
singularity ${\cal{O}}(\alpha^2)$ relativistic corrections to
the non-relativistic cross section of heavy fermion-antifermion pair
production in $e^+e^-$ annihilation cannot be
determined by means of conventional multi-loop perturbation theory.
\end{small}
\end{minipage}
\end{center}
\setcounter{footnote}{0}
\renewcommand{\thefootnote}{\arabic{footnote}}
\vspace{1.2cm}
\newpage
%
%
\noindent
\section{Introduction}
\label{sectionintroduction}
In view of future experiments (NLC, B-factory, $\tau$-charm factory) 
where heavy quark-antiquark pairs will
be produced in the kinematic region close to the threshold and a large
amount of data can be expected, it is a very attractive idea that an
extraction of the strong coupling $\alpha_s$ at a specific scale (or
equivalently $\Lambda_{\mbox{\tiny QCD}}$) might be possible which is
accurate enough to allow for a serious comparison to complementary
determinations of $\alpha_s$ from high energy experiments, where
quark masses are much smaller than the relevant energy scales. Such an
analysis would be an extremely important test of QCD. In recent
literature two attempts can be found~\cite{Voloshin1,Pich1} where such
an analysis has
been carried out based on present data on properties of $b\bar b$
mesons and on theoretical calculations involving well known results in
the non-relativistic limit. The results of these analyses
are somewhat controversial indicating that a better understanding of
the structure and size of relativistic corrections to the
non-relativistic limit and of the interplay of these corrections with
non-perturbative effects is mandatory.
\par
The framework in which relativistic corrections can be determined
systematically in a very elegant way is non-relativistic quantum
chromodynamics (NRQCD)~\cite{Caswell1} which is based on the concept of
effective field theories. NRQCD consists of a non-relativistic
Schr\"odinger field theory with a Coulomb-like QCD potential whereby
relativistic effects are incorporated by introduction of higher
dimensional operators in accordance to the underlying symmetries. In
order to render NRQCD equivalent to QCD the NRQCD Lagrangian has to
be matched to predictions in the framework of conventional multi-loop
perturbation theory. This procedure leads to, in general,
divergent renormalization constants multiplying the operators in the
NRQCD Lagrangian and is essentially equivalent to a separation of
short- and long-distance effects. As far as the decay and production
properties of a heavy quark-antiquark pair involving single photon
annihilation in the threshold regime are concerned the relevant parts
of the NRQCD Lagrangian have only been renormalized at leading and
next-to-leading order in $\alpha_s$ so far~\cite{Bodwin1}.
\par
In this letter we present the two-loop contributions to the
electromagnetic vertex describing the decay of a virtual photon into
two massive fermions in the kinematic regime where the squared photon
four momentum is close to four times the squared fermion mass. The
calculation is performed in the framework of QED where only one
fermion 
species with mass $M$ and electronic charge $e$ exists.
The result is presented up to next-to-next-to-leading order (NNLO)
in an expansion in 
\begin{equation}
\beta \, = \, \sqrt{1-4\,\frac{M^2}{q^2+i\epsilon}}
\,,
\end{equation}
which is equal to the velocity 
of the fermions in the c.m. frame above threshold\footnote{
Thus $\beta$ will be called ``velocity'' for the rest of this paper.
In this paper we use the notion ``leading order'' (and NLO, NNLO, NNNLO)
exclusively for the expansion in the velocity.
},
$\sqrt{q^2}$ being the c.m. energy.
We analyse the structure and form of the results and demonstrate the
existence of a new logarithmic Coulomb singularity occurring at NNLO
in the velocity expansion. In
particular, we will study the impact of this singularity on the massive
fermion-antifermion pair production cross section slightly above the
threshold. In the framework of QCD our two-loop results represent all
two-loop contributions involving the color factor $C_F^2$ (from
exchange of
two virtual gluons) and $C_F T$ (from the exchange of one gluon with
the insertion of the fermion-antifermion vacuum polarization)
and, therefore, are a  gauge invariant subset of all
two-loop QCD contributions in the threshold regime\footnote{
The two-loop contributions arising from the virtual effects
of massless fermions have been calculated in~\cite{Hoang1}
for all ratios $M^2/q^2$ above threshold
and will not be discussed in this work.
}. 
\par
The two-loop contributions calculated in this work represent a first
step toward a two-loop renormalization of the NRQCD Lagrangian
describing single photon annihilation processes involving heavy
quark-antiquark pairs. In particular, they are a crucial input for the
determination of NNLO relativistic corrections for the
single photon annihilation contributions to decay and production of
heavy quark-antiquark bound states and for the production of heavy
quark-antiquark 
pairs in $e^+e^-$ collisions slightly above threshold. In the
framework of QED the result is essential for the determination of the
single photon annihilation contributions to the ${\cal{O}}(\alpha^6)$
triplet-singlet hyperfine splitting of the
positronium ground state.
\par
The content of this work is organized as follows: in
Section~\ref{sectionnotation} we explain the notation and introduce
the electromagnetic form factors relevant for our calculations and
discussions. In Section~\ref{sectiononeloop} we reanalyse the
well-known one-loop contributions to the form factors in the threshold
region. We discuss the structure and properties of the individual
coefficients of the expansion in small $\beta$ and derive predictions
for the form of the two-loop corrections based on the factorization of
long- and short-distance contributions. In
Section~\ref{sectiontwoloop} the two-loop corrections are explicitly
calculated using the dispersion integration technique. It is
demonstrated that the predictions of Section~\ref{sectiononeloop} 
are realized and the logarithmic Coulomb singularity is
discussed. Section~\ref{sectionsummary} contains a summary. 

\vspace{1cm}
\noindent
\section{Notation and Definition of the Electromagnetic Form Factors}
\label{sectionnotation}
It is common to parameterize radiative (multi-loop) corrections to the
electromagnetic vertex, describing the decay of a photon with
virtuality $q^2$ into a fermion-antifermion pair, in terms of the Dirac 
($F_1$) and the Pauli ($F_2$) form factors. They are defined through
the relation
\begin{equation}
\bar{u}(p^\prime)\,\Lambda_\mu^{em}\,v(p) \,=\,
i\,e\,\bar{u}(p^\prime)\,\bigg[\,
\gamma_\mu\,F_1(q^2) 
+ \frac{i}{2\,M}\,\sigma_{\mu\nu}\,q^\nu\,F_2(q^2)
\,\bigg]\,v(p)
\,,
\end{equation}
where
\[ q \, = \, p+p^\prime
\hspace{.5cm}\mbox{and}\hspace{.5cm}
\sigma_{\mu\nu} \, = \, \frac{i}{2}\,[\,\gamma_\mu,\gamma_\nu\,]
\,.
\]
Expanded in the number of loops, which corresponds to an expansion in
powers of the fine structure constant $\alpha$, the form
factors  $F_1$ and $F_2$ read
\begin{eqnarray}
F_1(q^2) & = & 1 \, + \, \Big(\frac{\alpha}{\pi}\Big)\,F_1^{(1)}(q^2) 
\, + \,
\Big(\frac{\alpha}{\pi}\Big)^2\,F_1^{(2)}(q^2) \, + \, \cdots
\,,\nonumber\\[2mm]
F_2(q^2) & = & \hspace{0.8cm} \Big(\frac{\alpha}{\pi}\Big)\,
F_2^{(1)}(q^2) \, + \,
\Big(\frac{\alpha}{\pi}\Big)^2\,F_2^{(2)}(q^2) \, + \, \cdots
\,.
\end{eqnarray}
The use of $F_1$ and $F_2$ is particularly convenient for the
kinematic point $q^2=0$ because $F_2(0)=(g_f-2)/2$ is directly
related to the gyro-magnetic ratio of the fermion and because
$F_1(0)=1$ (i.e. $F_1^{(n)}(0)=0$ for $n=1,2,\ldots,\infty$)
due to gauge invariance. These properties are useful if dispersion
relation
techniques are used to calculate higher loop contributions because
overall UV divergences to $F_1^{(n)}$ ($n=1,2,\ldots,\infty$) can be
automatically renormalized by using once-subtracted dispersion
relations. For $F_2$, on the other hand, no overall UV divergences
exist which
makes the use of unsubtracted dispersion relations convenient. Since
the determination of our two-loop results relies on the dispersion
relation technique we will use the form factors $F_1$ and $F_2$ for
the actual calculations.
\par
For physical applications in the threshold region, where $q^2\approx 4
M^2$, however, the use of the combinations
\begin{eqnarray}
G_m & = & F_1 + F_2 \,,
\label{Gmdefinition} \\[2mm]
G_e & = & F_1 + \frac{s}{4\,M^2}\,F_2
\,.
\label{Gedefinition}
\end{eqnarray}
is more appropriate. This can be easily seen by considering the
contributions of the form factors $F_1$ and $F_2$ to the cross section
for the production of a fermion-antifermion pair (with fermion mass $M$) 
in $e^+e^-$ annihilation.
Taking the colliding electrons and positrons as
massless one arrives at the following angular distribution for the
produced fermion pairs for the c.m. energy $\sqrt{q^2}$ above threshold
\begin{equation}
 \frac{\rm d \,\sigma(e^+ e^- \to f\bar f)}{\rm d\,\Omega} \, = \,
  \frac{\alpha^2\,\beta}{4\,q^2}\,\bigg[\,  
   |G_m|^2\,(1+\cos^2\theta) +
   \frac{4\,M^2}{q^2}|G_e|^2\,\sin^2\theta \,\bigg]
\,,
\label{angulardistribution}
\end{equation}
where $\theta$ is the deflection angle. The corresponding expression
for the total cross section reads 
($\sigma_{pt} = 4 \pi \alpha^2 / 3 q^2$)
\begin{equation}
R \, \equiv \,
\frac{\sigma(e^+e^- \to f\bar f)}{\sigma_{pt}} \, = \,
\beta\,\bigg[\, |G_m|^2 + \frac{1}{2}\,(1-\beta^2)\,|G_e|^2
\,\bigg]
\,.
\label{totalcrosssection}
\end{equation} 
$G_m$ and $G_e$ are called magnetic and electric form factors,
respectively~\cite{Renard1}. They can be easily identified as the total
spin projection (relative to the electron direction) $\pm 1$ and $0$
amplitudes describing the produced fermion-antifermion pair 
in a triplet ($J^{PC}=1^{-\,-}$) state. Because the
fermion-antifermion production cross section represents one of the
most important applications of the corrections to the electromagnetic
vertex we will discuss the structure and
properties of the corrections by analysing the moduli squared of the
magnetic and electric form factors above threshold. Their expansion in
the number of loops (i.e. in powers of the fine structure constant)
reads 
\begin{eqnarray}
|G_m|^2 & = &
1 \, + \, \Big(\frac{\alpha}{\pi}\Big)\,g_m^{(1)}
\, + \,
\Big(\frac{\alpha}{\pi}\Big)^2\,g_m^{(2)}
\, + \, \cdots
\,,
\nonumber\\[2mm]
|G_e|^2 & = &
1 \, + \, \Big(\frac{\alpha}{\pi}\Big)\,g_e^{(1)}
\, + \,
\Big(\frac{\alpha}{\pi}\Big)^2\,g_e^{(2)}
\, + \, \cdots
\,.
\end{eqnarray}
We finally would like to emphasize that throughout this paper the
fermions are understood as stable particles and that the on-shell
renormalization scheme is employed, where $\alpha=1/137$ and $M$ is
the fermion pole mass. 

\vspace{1cm}
\section{One-Loop Results}
\label{sectiononeloop}
Analytic expressions for 
the one-loop contributions to the electromagnetic vertex valid for all
energies are well known since quite a long time~\cite{Kallensabry1,
Schwinger1}.  In this
section we reanalyse the one-loop contributions in the threshold
region in the velocity expansion as a preparation for the
examination of the two-loop contributions in
Section~\ref{sectiontwoloop}.
\par
Regularizing the soft photon infrared divergences with a fictitious
small photon mass $\lambda$, where the hierarchy $\lambda/M\ll|\beta|\ll
1$ is understood, the one-loop contributions to the electromagnetic
form factors $F_1$ and $F_2$ assume the form
\begin{eqnarray}
F_1^{(1)}(q^2) & \stackrel{\beta\to 0}{=} &
i\,\frac{\pi }{2\,\beta}\,\bigg[\, \ln\Big(-\frac{2\,i\,\beta\,M}{\lambda}\Big) - 
     \frac{1}{2} \,\bigg]  \,-\, \frac{3}{2} \,+\, 
  i\,\frac{\pi \,\beta}{2}\,\bigg[\, \ln\Big(-\frac{2\,i\,\beta\,M}{\lambda}\Big) - 
     \frac{1}{2} \,\bigg] \,\nonumber\\ 
 & & \mbox{} \, -\, 
  \frac{4}{3}\,\bigg[\, \ln\Big(\frac{M}{\lambda}\Big) + \frac{5}{24} \,\bigg] \,
   {{\beta}^2}
\, + \, {\cal{O}}(\beta^3)
\,,
\label{F1oneloop}
\\[2mm]
F_2^{(1)}(q^2) & \stackrel{\beta\to 0}{=} &
i\,\frac{\pi }{4\,\beta} \,-\, \frac{1}{2} \,-\, i\,\frac{\pi\,\beta}{4} \,+\, 
  \frac{1}{3}\,{{\beta}^2}
\, + \, {\cal{O}}(\beta^3)
\label{F2oneloop}
\end{eqnarray}
in the velocity expansion up to NNNLO. 
Expressions~(\ref{F1oneloop}) and
(\ref{F2oneloop}) are valid above as well as below the threshold
point, $q^2=4 M^2$, and lead to the following one-loop contributions
to the moduli squared of the magnetic and electric form factors above
the threshold
\begin{eqnarray}
\Big(\frac{\alpha}{\pi}\Big)\,g_m^{(1)}(q^2)
 & \stackrel{\beta\to 0}{=} &
\frac{\alpha\,{{\pi }}}{2\,\beta} - 4\,\frac{\alpha}{\pi} + 
\frac{\alpha\,{{\pi }}\,\beta}{2} - 
  \frac{\alpha}{3\,\pi}\,\bigg[\, 8\,\ln\Big(\frac{M}{\lambda}\Big) -
\frac{1}{3}
\,\bigg] \,{{\beta}^2}
\, + \, {\cal{O}}(\beta^3)
\,,
\label{gmoneloop}
\\[2mm]
\Big(\frac{\alpha}{\pi}\Big)\,g_e^{(1)}(q^2)
 & \stackrel{\beta\to 0}{=} &
\frac{\alpha\,{{\pi }}}{2\,\beta} - 4\,\frac{\alpha}{\pi} + 
\frac{\alpha\,{{\pi }}\,\beta}{2} - 
  \frac{8\,\alpha}{3\,\pi}\,\bigg[\, \ln\Big(\frac{M}{\lambda}\Big) +
\frac{1}{3} 
\,\bigg] \,{{\beta}^2}
\, + \, {\cal{O}}(\beta^3)
\,.
\label{geoneloop}
\end{eqnarray}
For the rest of this section we will discuss the individual terms in
the velocity expansion displayed in eqs.~(\ref{gmoneloop}) and 
(\ref{geoneloop}). We would
like to emphasize that most of the issues which are  mentioned are
well known and
have been noted before at various places throughout the
literature. However, 
we think that a review of these topics is necessary for a 
better understanding of the structure of the two-loop results
presented in Section~\ref{sectiontwoloop} and the new
information contained in them.
\par
Expressions~(\ref{gmoneloop}) and (\ref{geoneloop}) exhibit the well
known soft photon
divergence $\propto \ln(M/\lambda)$ which arises from the masslessness
of the photon. This divergence occurs at order $\beta^2$ and would
cancel
with the corresponding soft photon divergence coming from the process
of real radiation of one photon off one of the fermions according to
the Kinoshita-Lee-Nauenberg theorem~\cite{Kinoshita1,LeeNauenberg1}. 
The fact that the
divergent term  $\ln(M/\lambda)$ is suppressed by $\beta^3$ relative to
the leading contribution in the expansion in $\beta$ is expected at
any loop level because close to threshold the real radiation of one
photon results in an additional factor $\beta$ from the phase space
needed for the photon and a factor $\beta^2$ from the square of the
dipole matrix element\footnote{
It should be noted that this statement is equivalent to the fact that
contributions from the non-instantaneous (i.e. transverse)
exchange of photons among the  fermion-antifermion pair 
are suppressed by $\beta^3$ with respect to the leading contributions
in the velocity expansion. As an example, this feature is apparent
in a ${}^3S_1$, $J^{PC}=1^{-\,-}$ fermion-antifermion bound state,
where the velocity $\beta$ of the fermions is of order
$\alpha$. There,
the exchange of non-instantaneous photons leads to the Lamb shift
which represents a ${\cal{O}}(\alpha^3)$ correction
relative to the Coulomb energy levels. (See also~\cite{Grinstein1}.)
}.
Because the soft photon $\ln(M/\lambda)$ divergence indicates the
inadequacy of a pure fermion-antifermion final state and the need for
the introduction of a higher fock fermion-antifermion-photon state,
the $\beta^3$ suppression allows us to conclude that the notion of a
pure
fermion-antifermion state is consistent if we are only interested in
NNLO accuracy in the expansion in $\beta$.
\par
The leading term in the velocity expansion in eqs.~(\ref{gmoneloop})
and (\ref{geoneloop})
is the well known Coulomb singularity which diverges for $\beta\to
0$. Similar to the soft photon divergence discussed above the Coulomb
singularity arises from the fact that the photon is massless and 
represents a long-distance effect. The Coulomb singularity, however, 
is of completely different nature. Whereas the soft photon singularity
indicates the inadequacy of a pure fermion-antifermion state beyond
NNLO in the velocity expansion the Coulomb singularity reveals that in
the non-relativistic limit (corresponding to the leading order in the
velocity expansion) the photon-mediated interaction between the
fermion-antifermion pair cannot be described in an expansion in
Feynman diagrams, where a diagram with a larger number of loops
(corresponding to a larger number of exchanged photons) would
represent a higher order correction. Rather, a resummation of diagrams
with any number of exchanged photons is needed to arrive at a sensible
description of the interaction between the fermion-antifermion
pair. The leading contribution in the velocity expansion is obtained
by resummation of diagrams with instantaneous Coulomb exchanges of
longitudinal photons (in the Coulomb gauge). This procedure can be
explicitly carried out by calculating the normalized
wave function at the 
origin, $\Psi_{E}(0)$, to the Schr\"odinger equation describing a 
non-relativistic fermion-antifermion pair with a Coulomb interaction 
potential for positive energies $E=M \beta^2$. The result of this 
calculation reads (see e.g.~\cite{Schwinger1, Sakharov1, Fadin1})
\begin{equation}
|G_m|^2_{\mbox{\tiny{LO}}} \, = \,
|G_e|^2_{\mbox{\tiny{LO}}} \, = \,
|\Psi_{M \beta^2}(0)|^2 
\, = \,
\frac{z}{1-\exp{(-z)}}
\label{LOsommerfeld}
\,,
\end{equation}
where
\begin{equation}
z \, \equiv \, \frac{\alpha\,\pi}{\beta}
\,,
\end{equation}
and is often called ``Sommerfeld factor'' in the literature.
The $1/\beta$ Coulomb singularity in eqs.~(\ref{gmoneloop})
and (\ref{geoneloop})
can be recovered as the ${\cal{O}}(\alpha)$ contribution in the
expansion of the Sommerfeld factor for $\alpha\ll\beta$,
\begin{equation}
\frac{z}{1-\exp{(-z)}}
\, \stackrel{\alpha\ll\beta}{=} \,
1 \,+\, \frac{z}{2} \,+\, \frac{z^2}{12} \,+\, {\cal{O}}(\alpha^3)
\,.
\label{sommerfeldexpansion}
\end{equation}
This, on the
other hand, also shows that the velocity expansion of the
perturbative (in the number of loops) series can only be
applied in the limit $\alpha\ll\beta\ll 1$, where an expansion
in the number of loops (i.e. in $\alpha$) is justified\footnote{
It should be noted that the region of convergence of 
the Taylor expansion
\[
\frac{z}{1-\exp{(-z)}} \, = \,
1+\frac{z}{2}+
\sum\limits_{n=1}^{\infty}
(-1)^{n+1}\,\frac{B_n\,z^{2n}}{(2n)!}
\,,
\]
where $B_n$ are the Bernoulli numbers ($B_1=1/6$, $B_2=1/30$,
$B_3=1/42\,,\ldots$),
is $|z|<2\pi \, \Leftrightarrow \, |\beta| >
\alpha/2$. This shows that for phenomenological applications
a resummation of the leading order contributions in the velocity
expansion to any number of loops
is mandatory in the kinematic regime $|\beta|\lsim\alpha$.
}.
It is worth to study the effect of this resummation: inserting
the Sommerfeld factor into the formula for the cross 
section, eq.~(\ref{totalcrosssection}), we get at threshold
\begin{equation}
R\,\sim\,\frac{3}{2}\,\beta\,\frac{z}{1-\exp{(-z)}} 
\, \stackrel{\beta\to 0}{\longrightarrow}  \, 
 \frac{3}{2}\,\alpha\,\pi \,,
\label{bto0correct}
\end{equation}
which is the correct result according to non-relativistic
quantum mechanics.
On the other hand, if we naively use the one-loop result 
(i.e. expansion in small $\alpha$), we obtain
\begin{equation}
R\,\sim\,\frac{3}{2}\,\beta\,(1+\frac{z}{2})
\, \stackrel{\beta\to 0}{\longrightarrow}\,   
 \frac{3}{4}\,\alpha\,\pi\,.
\end{equation}
Clearly, the perturbative calculation in the number of loops, which is
based on the assumption that $\alpha$ is a valid expansion parameter
close to threshold, gives a prediction for $R$ at threshold which 
deviates from the correct one by a factor of one half. 
\par
The next-to leading contribution in the velocity expansion in
eqs.~(\ref{gmoneloop}) and (\ref{geoneloop}), $-4\,\alpha/\pi$,
represents a short-distance
correction and can be understood as a finite ${\cal{O}}(\alpha)$
renormalization of the electromagnetic current which produces the
fermion-antifermion pair in the threshold
region. The short-distance character of this ${\cal{O}}(\alpha)$
correction has been demonstrated explicitly by the calculation of 
the BLM (Brodsky-Lepage-Mackenzie~\cite{Brodsky1}) 
scale in the coupling governing the $-4\, \alpha/\pi$
contribution~\cite{Hoang1,Hoang2,Voloshin1}.
This BLM scale is of order the fermion mass $M$ and indicates that
the $-4\, \alpha/\pi$ contribution represents a correction
to the fermion-antifermion production process which occurs at short
distances of order $1/M$. In contrast, the BLM scale of the coupling 
in the leading
term in the velocity expansion, $\alpha \pi/2 \beta$, is of order
of the relative momentum of the fermion-antifermion pair,
$M \beta$~\cite{Hoang1,Hoang2}, indicating that the latter
contribution belongs
to the fermion-antifermion wave function. 
As a consequence the leading order (long-distance) contributions
contained in the Sommerfeld factor and the short-distance corrections
are expected to factorize which leads to
\begin{equation}
|G_m|^2_{\mbox{\tiny{NLO}}} \, = \,
|G_e|^2_{\mbox{\tiny{NLO}}} \, = \,
\frac{z}{1-\exp{(-z)}}
\,\bigg(\, 1-4\,\frac{\alpha}{\pi}
\,\bigg)
\label{NLOsommerfeld}
\end{equation}
for the NLO expressions in the velocity expansion of the moduli 
squared of the
magnetic and electric form factors in the threshold region. It should
be noted that the factorized result~(\ref{NLOsommerfeld}) resums all
contributions $(\alpha/\beta)^n\times[1,\alpha]$,
$n=0,1,2,\ldots,\infty$. Because no $(\alpha/\beta)^n \beta$
contributions exist\footnote{
Pure $\beta$-dependent corrections to the Sommerfeld factor are
of kinematic origin and therefore expected to be of NNLO in the 
velocity expansion, i.e. $\propto (\alpha/\beta)^n \beta^2$,
$n=0,1,2,\ldots,\infty$.
},
expression~(\ref{NLOsommerfeld}) unambiguously
predicts the leading and next-to-leading order contributions in the
velocity expansion for all $g_{m/e}^{(n)}$, $n=2,3,\ldots,\infty$. 
\par
The NNLO term in the velocity expansion in eqs.~(\ref{gmoneloop})
and (\ref{geoneloop}),
$\alpha \pi \beta/2$, has not received much attention in the
literature so far. Its structure, which involved the same power of
$\pi$ and the same coefficient $1/2$ as the LO term in the velocity
expansion,
strongly implies that it is of long-distance origin and therefore
belongs to the Sommerfeld factor. This is in accordance to the
observation that the BLM scale in the coupling of the term 
$\alpha \pi \beta/2$ is of order $M \beta$ rather than
$M$~\cite{Hoang1}. The relativistic extension of the Sommerfeld factor
(including ${\cal{O}}(\beta^2)$ corrections) should then read
\begin{equation}
\frac{\tilde z}{1-\exp(-\tilde z)} 
\,,\qquad
\tilde z\equiv \frac{\alpha\,\pi}{\beta}\,(1+\beta^2)
\,.
\label{Ob2sommerfeld}
\end{equation}
Although the arguments given above in favor of
expression~(\ref{Ob2sommerfeld}) are far from being a strict proof the
form of $\tilde z$  is very
convincing because it indicates that the relativistic relative
velocity $v_{rel}$ of the fermion-antifermion pair in the c.m. frame
is involved in the argument of the Sommerfeld factor if
${\cal{O}}(\beta^2)$ relativistic corrections are taken into account,
\begin{equation}
\tilde z \, = \, \frac{2\,\alpha\,\pi}{v_{rel}} 
\,,\qquad
v_{rel} \, = \, \frac{2\,\beta}{1+\beta^2}
\,.
\label{vreldef}
\end{equation}
Combining expression~(\ref{Ob2sommerfeld}) with the short-distance
factor $(1-4\alpha/\pi)$ and taking into account that no soft
photon divergence $\propto \ln(M/\lambda)$ arises up to NNLO in the
velocity expansion we can now predict that the two-loop
contributions to $|G_{m/e}|^2$ must have the form
\begin{equation}
g_{m/e}^{(2)}(q^2)
\, \stackrel{\beta\to 0}{=} \,
\frac{\pi^4}{12\,\beta^2} \, - \,
2\,\frac{\pi^2}{\beta} \, + \,
\frac{\pi^4}{6} \, + \,
\bigg[\, \mbox{finite terms without $\pi^4$}
\,\bigg] 
\, + \,
{\cal{O}}(\beta)
\,.
\label{gtwolooppredict}
\end{equation}
We want to emphasize the the ${\cal{O}}(1/\beta^2)$, 
${\cal{O}}(1/\beta)$ and ${\cal{O}}(\beta^0 \pi^4)$ contributions on
the r.h.s. of eq.~(\ref{gtwolooppredict}) are an unambiguous
prediction and have to be recovered in the explicit two-loop result if
the concept of factorization in the threshold regime is valid.
It should be noted that up to NNLO in the velocity expansion only the
${\cal{O}}(\beta^0)$ contributions symbolized by $[\mbox{finite terms
without $\pi^4$}]$ contain new two-loop information.

\vspace{1cm}
\noindent
\section{Two-Loop Results}
\label{sectiontwoloop}
To determine the two-loop contributions to the electromagnetic form
factors $F_1$ and $F_2$ in the velocity expansion we use the
dispersion integration technique. For that we have to integrate over
the absorptive parts $\mbox{Im} F_{1/2}^{(2)}$ which have been
determined 
a long time ago by Barbieri, Mignaco and Remiddi~\cite{Barbieri1},
\begin{eqnarray}
F_1^{(2)}(q^2) & = & 
-\,\frac{4\,M^2\,q^2}{q^2-4\,M^2}\,F_1^{\prime\,(2)}(0)
\nonumber\\ & &
 \, + \,\frac{1}{\pi}\,\frac{q^4}{q^2-4\,M^2}\,
\int\limits_{4\,M^2}^\infty
\frac{d q^{\prime\,2}}{ q^{\prime\,2}(q^{\prime\,2}-q^2-i\epsilon) }\,
\frac{q^{\prime\,2}-4\,M^2}{q^{\prime\,2}}\,
\mbox{Im}F_1^{(2)}(q^{\prime\,2})
\,,
\label{dispersionF1}
\\[3mm]
F_2^{(2)}(q^2) & = & 
-\,\frac{4\,M^2}{q^2-4\,M^2}\,F_2^{(2)}(0) 
\nonumber\\ & &
\, + \,\frac{1}{\pi}\,\frac{q^2}{q^2-4\,M^2}\,
\int\limits_{4\,M^2}^\infty
\frac{d q^{\prime\,2}}{q^{\prime\,2}-q^2-i\epsilon}\,
\frac{q^{\prime\,2}-4\,M^2}{q^{\prime\,2}}\,
\mbox{Im}F_2^{(2)}(q^{\prime\,2})
\,.
\label{dispersionF2}
\end{eqnarray}
We would like to mention that relations~(\ref{dispersionF1}) and
(\ref{dispersionF2}) are equivalent to the common once-subtracted and
unsubtracted dispersion relations. We use~(\ref{dispersionF1}) and 
(\ref{dispersionF2}) because they do not
run into non-analyticity problems in the integration region where
${q^\prime}^2-4 M^2$ is of order $\lambda^2$ if the limit $\lambda\to
0$ is
already taken before the integration. Since the absorptive parts
in~\cite{Barbieri1} are given in exactly this
limit~(\ref{dispersionF1}) and
(\ref{dispersionF2}) are more convenient because in them the integration
regime ${q^\prime}^2-4 M^2\approx\lambda^2$ is strongly
suppressed. The (low) price one has to pay is that the
${\cal{O}}(\alpha^2)$ fermion charge
radius~\cite{Appelquist1,Barbieri2},  
\begin{equation}
F_1^{\prime\,(2)}(0) \, = \,
\frac{1}{M^2}\,\bigg[\,
\frac{{{\pi }^2}}{6}\,\bigg(\, 3\,\ln 2 - \frac{49}{72} \,\bigg)  - 
  \frac{3}{4}\,\zeta_3 - \frac{4819}{5184}
\,\bigg]
\,,
\end{equation}
and the
${\cal{O}}(\alpha^2)$ anomalous magnetic
moment~\cite{Sommerfeld1,Petermann1}, 
\begin{equation}
F_2^{(2)}(0) \, = \,
\frac{{{\pi }^2}}{12}\,\bigg(- 6\,\ln 2 +1\,\bigg)  +
\frac{3}{4}\,\zeta_3 + 
  \frac{197}{144}
\,,
\end{equation}
have to be taken as an input\footnote{
This fact has already been pointed out in~\cite{Barbieri1}. We also
refer the
reader to this reference for a more thorough discussion of the
problems which occur in the integration region ${q^\prime}^2-4
M^2\approx \lambda^2$.
}.
Details for the quite lengthy but straightforward calculation of the
integrals~(\ref{dispersionF1}) and (\ref{dispersionF2}), which requires
strong support of algebraic manipulation programs, shall be
presented elsewhere.
\par
The final results for the two-loop contributions to $F_1$ and $F_2$ up
to NNLO in the velocity expansion read
\begin{eqnarray}
F^{(2)}_{1,2\gamma}
& \stackrel{\beta\to 0}{=} &
-\frac{\pi^2}{8\,{{\beta}^2}}\,\bigg[\, \frac{{{\pi }^2}}{6} + 
     \bigg(\, {{\ell}^2} - \ell + \frac{1}{3} \,\bigg)  
\,\bigg]  + 
  i\,\frac{\pi }{4\,\beta}\,\bigg[ -3\,\ell + 1 \,\bigg] \,
\label{F1twoloop2gamma}   \\ 
 & & \mbox{} - \bigg[\, \frac{{{\pi }^4}}{24} + 
    \frac{{{\pi }^2}}{4}\,\bigg(\, {{\ell}^2} - \ell + 
       \frac{23}{15}\,\ln(-i\,\beta) + \frac{7}{10}\,\ln 2 + 
       \frac{73}{50} \,\bigg)  + 
    \frac{9}{80}\,\left( 9\,\zeta_3 - \frac{421}{27} \right)  \,\bigg]
\,+\, {\cal{O}}(\beta) 
\,,
\nonumber
\\[2mm]
F^{(2)}_{1,f}
& \stackrel{\beta\to 0}{=} &
-\frac{13\,{{\pi }^2}}{45} + \frac{37}{12}
\,+\, {\cal{O}}(\beta^2) 
\,,
\label{F1twoloopf}
\\[2mm]
F^{(2)}_{2,2\gamma} 
& \stackrel{\beta\to 0}{=} &
-\frac{{{\pi }^2}}{8\,{{\beta}^2}}\,
     \bigg[\, \ell - {1\over 3} \,\bigg]   - 
  i\,\frac{\pi }{4\,\beta}\,\bigg[\, \ell + 1 \,\bigg] \,\nonumber\,
   \\ 
 & & \mbox{} + \bigg[\,\frac{{{\pi }^2}}{20}\,
   \bigg(\, \ln(-i\,\beta) + \frac{101}{6}\,\ln 2 - 
     \frac{559}{45} \,\bigg)  + \frac{1}{80}\,
   \bigg(\, 41\,\zeta_3 + \frac{269}{3} \,\bigg) \,\bigg]
\,+\, {\cal{O}}(\beta) 
\,,
\label{F2twoloop2gamma}
\\[2mm]
F^{(2)}_{2,f} 
& \stackrel{\beta\to 0}{=} &
\frac{{{\pi }^2}}{15} - \frac{23}{36}
\,+\, {\cal{O}}(\beta^2) 
\,,
\label{F2twoloopf}
\end{eqnarray}
where
\begin{equation}
\ell \, \equiv \, \ln\Big(-\frac{2\,i\,\beta\,M}{\lambda}\Big)
\end{equation}
and, as in the one-loop case, the hierarchy $\lambda/M\ll|\beta|\ll 1$
is understood.
In eqs.~(\ref{F1twoloop2gamma})--(\ref{F2twoloopf}) the contributions
from diagrams with two photons (subscript $2 \gamma$) and from the
diagrams with one photon
and the insertion of the fermion-antifermion vacuum
polarization\footnote{
The two-loop corrections $F^{(2)}_{1,f}$ and $F^{(2)}_{2,f}$ have
already been calculated before in~\cite{Hoang1} for all energies above 
threshold.
} 
(subscript $f$)
are displayed separately. This will facilitate the application in the
framework of QCD where both types of contributions are multiplied by
the different color
factors $C_F^2$ and $C_F T$, respectively, and represent gauge invariant
subsets of the full QCD two-loop contributions.
\par
The results~(\ref{F1twoloop2gamma})--(\ref{F2twoloopf}) lead to the
following two-loop contributions to the moduli squared of the
magnetic and electric form factors above threshold up to
NNLO in the velocity expansion
\begin{eqnarray}
g_m^{(2)}(q^2) & \stackrel{\beta\to 0}{=} &
\frac{{{\pi }^4}}{12\,{{\beta}^2}} - 2\,\frac{{{\pi }^2}}{\beta} + 
  \frac{{{\pi }^4}}{6} + {{\pi }^2}\,
   \bigg( -\frac{2}{3}\,\ln\beta   + \frac{4}{3}\,\ln 2 - 
     \frac{29}{12} \,\bigg)  - \zeta_3 + \frac{527}{36}
\,+\, {\cal{O}}(\beta) 
\,,
\label{gmtwoloop}
\\[2mm]
g_e^{(2)}(q^2) & \stackrel{\beta\to 0}{=} &
\frac{{{\pi }^4}}{12\,{{\beta}^2}} - 2\,\frac{{{\pi }^2}}{\beta} + 
  \frac{{{\pi }^4}}{6} + {{\pi }^2}\,
   \bigg( -\frac{2}{3}\,\ln\beta  + \frac{4}{3}\,\ln 2 - 
     \frac{7}{3} \,\bigg)  - \zeta_3 + \frac{527}{36}
\,+\, {\cal{O}}(\beta) 
\,.
\label{getwoloop}
\end{eqnarray} 
It is evident that the prediction made in the previous section based
on the one-loop corrections and on the factorization of long- and
short-distance contributions (see eq.~(\ref{gtwolooppredict})) are
indeed realized by our explicit two-loop result confirming the
statements given in Section~\ref{sectiononeloop}. As a consequence
only the ${\cal{O}}(\beta^0)$ terms in eqs.~(\ref{gmtwoloop}) and
(\ref{getwoloop}) essentially contain new information.
\par
The most conspicuous feature of the ${\cal{O}}(\beta^0)$ contributions
in eqs.~(\ref{gmtwoloop}) and (\ref{getwoloop}) is the term
$\ln(\beta)$. Similar to the $1/\beta^2$ Coulomb singularity exhibited
in the leading term in the velocity expansion, it indicates the
breakdown of the conventional perturbation series in the number of
loops in the limit 
$\beta\to 0$. The existence of this logarithm can be understood from
the fact that two scales are involved in the kinematic regime near
threshold, the fermion mass $M$ and the three momentum of the
fermion and antifermion in the c.m. frame $p\equiv M \beta$. The
logarithm of the velocity $\beta$ is therefore actually the logarithm
of the ratio of these two scales, $\ln(p/M)$. Because the soft scale
$p$ is characteristic for the fermion-antifermion wave function and
not relevant for the production mechanism of the fermion-antifermion
pair (which involves only the hard scale $M$), the $\alpha^2\ln(p/M)$
term in eqs.~(\ref{gmtwoloop}) and (\ref{getwoloop}) should occur 
with the same coefficient in the
${\cal{O}}(\alpha^2)$ corrections to the positronium decay rates. For
a viable comparison, however, we also have to include the
fermion-antifermion vacuum polarization effects coming from the fact
that the fermion-antifermion pair, which is in a
$J^{PC}=1^{-\,-}$ state, can virtually annihilate
into one photon. This can be easily achieved by  multiplying
$|G_{m/e}|^2$ by the
factor $|1+\Pi|^{-2}$, where $\Pi$ is the one-particle-irreducible
vacuum polarization function. The two-loop contribution to $\Pi$ also
contains a logarithm of $\beta$ in the velocity
expansion~\cite{Hoang3}. This leads to the additional contribution
$\alpha^2 \ln(\beta)$ which has to added to $-2\, \alpha^2
\ln(\beta)/3$ from $|G_{m/e}|^2$.
(Actually the spin average of the logarithmic terms in 
expressions~(\ref{gmtwoloop}) and (\ref{getwoloop}) has to be taken.
This trivially results in $-2\, \alpha^2 \ln(\beta)/3$
because the logarithmic term is universal in both spin 
amplitudes.)
Because the relative momentum of the electron-positron
pair in the positronium is of order $M \alpha$ we can expect that the
${\cal{O}}(\alpha^2)$ corrections to the (${}^3S_1$,
$J^{PC}=1^{-\,-}$) orthopositronium
decay rate should contain the contribution $\alpha^2 \ln(\alpha)/3$.
This logarithmic ${\cal{O}}(\alpha^2)$ correction has indeed been
found by explicit calculations of higher order correction to 
the orthopositronium decay
rate~\cite{Caswell2}. We therefore have to conclude that the
$\ln(\beta)$ term
in eqs.~(\ref{gmtwoloop}) and (\ref{getwoloop}) represents a new type
of Coulomb singularity which, similar to the power-like $1/\beta^n$
singularities, requires a resummation of contributions to all orders 
in the number of loops\footnote{
At this point we would like to mention that the logarithmic Coulomb
singularity has also been disussed 
in~\cite{Barbieri3} in the framework of quarkonia decays.
However, it is argued in~\cite{Barbieri3} (and also in~\cite{Bodwin1})
that this singularity (called ``logarithmic infrared divergence''
in~\cite{Bodwin1}) would indicate that perturbative QCD could not be
applied in the kinematic regime close to the threshold. We disagree
with this conclusion, because we think that this singularity can be
treated by a proper resummation of contributions to all orders in the
number of loops.
}. 
How such a resummation has to be carried out
for the $\ln(\beta)$ term in the vacuum polarization has been
demonstrated in~\cite{Hoang3}. 
\par
Finally, we want to discuss the impact of the $\ln(\beta)$ singularity
on the cross section of fermion-antifermion production very close to
threshold, see eqs.~(\ref{angulardistribution}) and
(\ref{totalcrosssection}). Because the moduli squared of the magnetic
and electric form factors are multiplied by the phase space factor
$\beta$ one might naively think that the $\ln(\beta)$ singularity is
suppressed by $\beta$ and does not affect the cross section for
$\beta\to 0$. At this point we have to emphasize that the same would
then be true for the  short-distance correction, 
$-4\,\alpha/\pi$, in the one-loop contribution to $|G_{m/e}|^2$
because the latter also represents a ${\cal{O}}(\beta^0)$ term in the
velocity expansion (see eqs.~(\ref{gmoneloop}) and (\ref{geoneloop})). 
However, the one-loop short-distance 
correction survives for $\beta\to 0$, see eqs.~(\ref{bto0correct}) and
(\ref{NLOsommerfeld}). The resolution of this apparent contradiction
comes from the fact that due to factorization (see
eq.~(\ref{NLOsommerfeld})) the one-loop short-distance
correction is also contained in the ${\cal{O}}(1/\beta)$ term of the
two-loop contribution to $|G_{m/e}|^2$ where it multiplies the
${\cal{O}}(\alpha)$ contribution of the expansion of the Sommerfeld
factor for small $\alpha$. This contribution does not vanish in the
cross section for $\beta\to 0$ and illustrates the mechanism why the
one-loop short-distance correction survives in this
limit. In order to see that something similar happens to the
$\ln(\beta)$ singularity in the two-loop results~(\ref{gmtwoloop}) and
(\ref{getwoloop}) let us have a closer look on the structure of the
one- and two-loop contributions to the form factors $F_1$ and $F_2$. It
has been shown by Yennie, Frautschi and Suura~\cite{Yennie1} that the
infrared soft photon divergences exponentiate completely. 
Because real soft photon divergences
in $|G_{m/e}|^2$ occur only beyond NNLO in the velocity expansion (see
Section~\ref{sectiononeloop}) all soft photon divergences which arise
up to NNLO in the velocity expansion in 
eqs.~(\ref{F1twoloop2gamma})--(\ref{F2twoloopf}) 
can be factorized into a divergent phase factor
which is known as the {\it Coulomb phase}. In the moduli squared of
the form factors this phase drops out. Since the Coulomb phase has
to be considered as an intrinsic property of the fermion-antifermion
wave function, where the relative momentum $2 M \beta$ is a relevant
scale, we can expect that the divergent phase factor should involve 
the logarithm of the ratio $2M\beta/\lambda$. This feature is indeed
realized because the sum of Born, one-loop and two-loop contributions
to $F_1$ and $F_2$ above threshold can be rewritten as
\begin{eqnarray}
\lefteqn{
1 \, + \, \Big(\frac{\alpha}{\pi}\Big)\,F_1^{(1)}
\, + \,
\Big(\frac{\alpha}{\pi}\Big)^2\,\bigg[\,
F_{1, 2\gamma}^{(2)} + F_{1, f}^{(2)}
\,\bigg]
}\nonumber\\ & \longrightarrow &
\exp\bigg\{\,i\,\frac{\alpha}{2} \,\bigg(\, \frac{1}{\beta} + 
   \beta \,\bigg) \, \ell\,\bigg\}\,
\,\bigg\{\,
 1 - \Big(\frac{\alpha}{\pi}\Big)\,\bigg[\, i\,\frac{\pi }{4}\,
        \left( \frac{1}{\beta} + \beta \right)  + {3\over 2} \,\bigg] 
 +    \Big(\frac{\alpha}{\pi}\Big)^2\,\bigg[\,  -\frac{{{\pi }^2}}
        {24\,{{\beta}^2}}\,
          \bigg(\, \frac{{{\pi }^2}}{2} + 1 \,\bigg)
\,\nonumber\,\\ 
 & & \mbox{}\,\qquad   + 
       i\,\frac{\pi }{4\,\beta}
        - \frac{{{\pi }^4}}{24} - 
       \frac{{{\pi }^2}}{20}\,\bigg(\, \frac{23}{3}\,
           \ln(-i\,\beta) + \frac{7}{2}\,\ln 2 + 
          \frac{1177}{90} \,\bigg)  - 
       \frac{9}{80}\,\bigg(\, 9\,\zeta_3 - 43 \,\bigg)  \,\bigg]
\,\bigg\}
\,,
\label{F1factorized}
\\[2mm] 
\lefteqn{
\Big(\frac{\alpha}{\pi}\Big)\,F_2^{(1)}
\, + \,
\Big(\frac{\alpha}{\pi}\Big)^2\,\bigg[\,
F_{2, 2\gamma}^{(2)} + F_{2, f}^{(2)}
\,\bigg]
}\nonumber\\ & \longrightarrow &
\exp\bigg\{\,i\,\frac{\alpha}{2}\,\bigg(\, \frac{1}{\beta} + \beta 
   \,\bigg) \, \ell \,\bigg\}
\,\bigg\{\, \Big(\frac{\alpha}{\pi}\Big)\,\bigg[\, i\,\frac{\pi }{4}\,
        \bigg(\, \frac{1}{\beta} - \beta \,\bigg)  - \frac{1}{2} \,\bigg]  + 
 \Big(\frac{\alpha}{\pi}\Big)^2\,\bigg[\, \frac{{{\pi }^2}}{24\,{{\beta}^2}}\,
        \nonumber\,\\ 
 & & \mbox{}\,\qquad   - 
       i\,\frac{\pi }{4\,\beta} + 
       \frac{{{\pi }^2}}{20}\,\bigg(\, \ln(-i\,\beta) + 
          \frac{101}{6}\,\ln 2 - \frac{499}{45} \,\bigg)  + 
       \frac{1}{80}\,\bigg(\, 41\,\zeta_3 + \frac{347}{9} \,\bigg) \,\bigg] 
\,\bigg\}
\,.
\label{F2factorized}
\end{eqnarray}
The factorized expressions~(\ref{F1factorized}) and
(\ref{F2factorized}) predict that at the three-loop level the real
parts of the form factors $F_1$ and $F_2$ contain the logarithmic 
and $\lambda$-independent ${\cal{O}}(1/\beta)$ contributions
$-23\, \alpha^3 \pi \ln(\beta)/ 240\beta$ and $\alpha^3 \pi
\ln(\beta)/ 80\beta$, respectively, in the velocity expansion
above the threshold. 
As a consequence, $|G_{m}|^2$ and $|G_{e}|^2$ both 
contain the three-loop term $-\alpha^3 \pi \ln(\beta) / 3\beta $ 
in the velocity expansion. We would like to
emphasize that the argument just given cannot be used to determine all
three-loop contributions, but it clearly shows that a logarithmic
Coulomb singularity also exists at order $\alpha^3/\beta$ which does
not vanish in the limit $\beta\to 0$ in the cross section. The
coefficient of this singularity further strongly implies that the 
$\ln(\beta)$ contributions in $|G_{m}|^2$ and $|G_{e}|^2$ to any
number of loops and at NNLO in the velocity  expansion above threshold
can be cast into the factorized form
\begin{equation}
\bigg[\,|G_{m/e}|^2\,\bigg]_
{\mbox{\tiny NNLO}\,\ln\beta-\mbox{\tiny contributions}} 
\, \sim \,
\frac{z}{1-\exp{(-z)}}
\,\bigg( - \frac{2}{3}\,\alpha^2\,\ln\beta
\,\bigg)
\,.
\label{Gmefactorizedthreeloop}
\end{equation}
It is clear from expression~(\ref{Gmefactorizedthreeloop}) and the
arguments given above that the logarithmic
Coulomb singularity does indeed affect the prediction for the cross
section for $\beta\to 0$. In particular, we conclude that a
conventional fixed order multi-loop 
calculation is not capable to determine NNLO
(i.e. ${\cal{O}}(\alpha^2)$)
relativistic corrections to the non-relativistic cross 
section\footnote{
In a recent publication where large-n QCD sum rules were
applied to the $b\bar b$ system~\cite{Pich1} it was claimed that 
${\cal{O}}(\alpha_s^2)$ accuracy
was achieved in the determination of the strong coupling
$\alpha_s$ and the bottom mass because two-loop corrections
to the cross section were taken into account. Because the large-n
limit peels out the threshold behavior of the $b\bar b$ production
cross section, the results presented in~\cite{Pich1} do not include 
NNLO relativistic effects properly and, therefore, are not at the
${\cal{O}}(\alpha_s^2)$ accuracy level. (See also~\cite{Hoang3}.)
}.
In order to determine the correct form of the NNLO relativistic
corrections to the non-relativistic cross section (or the Sommerfeld
factor) resummations of the type mentioned before have to be performed.
Such a program is beyond the scope of this work and will be carried
out elsewhere.

\vspace{1cm}
\noindent
\section{Summary}
\label{sectionsummary}
In this work we have determined the two-loop contributions to the
electromagnetic form factors in the kinematic regime close to the
fermion-antifermion threshold up to NNLO in an expansion in the
velocity of the fermions in the c.m.~frame. In the framework of NRQCD
and NRQED the results are an important input for the
two-loop renormalization of the effective Lagrangian.
As the main outcome of this work we have demonstrated the existence of
a new logarithmic (in the velocity) Coulomb singularity at NNLO in the
velocity expansion. This logarithmic contribution belongs to the
fermion-antifermion wave function and exists for the production of
free fermion-antifermion pairs above threshold as well as for
fermion-antifermion pairs in a bound state. For the case of
fermion-antifermion pair production in $e^+e^-$ annihilation the
logarithm indicates that a resummation of contributions to any number
of loops is mandatory in order to arrive at a viable (i.e. finite)
prediction for the cross section with NNLO accuracy very close to the
threshold point.

\par
\vspace{.5cm}
\section*{Acknowledgement}
I am grateful to J.H.~K\"uhn for suggesting this project.
I thank D.~Broadhurst, P.~Labelle and K.~Schilcher for useful
conversation. This work
is supported in part by the Department of Energy under contract
DOE~DE-FG03-90ER40546.

\vspace{1.0cm}
%
%
\sloppy
\raggedright
\def\app#1#2#3{{\it Act. Phys. Pol. }{\bf B #1} (#2) #3}
\def\apa#1#2#3{{\it Act. Phys. Austr.}{\bf #1} (#2) #3}
\def\lhc{Proc. LHC Workshop, CERN 90-10}
\def\npb#1#2#3{{\it Nucl. Phys. }{\bf B #1} (#2) #3}
\def\nP#1#2#3{{\it Nucl. Phys. }{\bf #1} (#2) #3}
\def\plb#1#2#3{{\it Phys. Lett. }{\bf B #1} (#2) #3}
\def\prd#1#2#3{{\it Phys. Rev. }{\bf D #1} (#2) #3}
\def\pra#1#2#3{{\it Phys. Rev. }{\bf A #1} (#2) #3}
\def\pR#1#2#3{{\it Phys. Rev. }{\bf #1} (#2) #3}
\def\prl#1#2#3{{\it Phys. Rev. Lett. }{\bf #1} (#2) #3}
\def\prc#1#2#3{{\it Phys. Reports }{\bf #1} (#2) #3}
\def\cpc#1#2#3{{\it Comp. Phys. Commun. }{\bf #1} (#2) #3}
\def\nim#1#2#3{{\it Nucl. Inst. Meth. }{\bf #1} (#2) #3}
\def\pr#1#2#3{{\it Phys. Reports }{\bf #1} (#2) #3}
\def\sovnp#1#2#3{{\it Sov. J. Nucl. Phys. }{\bf #1} (#2) #3}
\def\sovpJ#1#2#3{{\it Sov. Phys. LETP Lett. }{\bf #1} (#2) #3}
\def\jl#1#2#3{{\it JETP Lett. }{\bf #1} (#2) #3}
\def\jet#1#2#3{{\it JETP Lett. }{\bf #1} (#2) #3}
\def\zpc#1#2#3{{\it Z. Phys. }{\bf C #1} (#2) #3}
\def\ptp#1#2#3{{\it Prog.~Theor.~Phys.~}{\bf #1} (#2) #3}
\def\nca#1#2#3{{\it Nuovo~Cim.~}{\bf #1A} (#2) #3}
\def\ap#1#2#3{{\it Ann. Phys. }{\bf #1} (#2) #3}
\def\hpa#1#2#3{{\it Helv. Phys. Acta }{\bf #1} (#2) #3}
\def\ijmpA#1#2#3{{\it Int. J. Mod. Phys. }{\bf A #1} (#2) #3}
\def\ZETF#1#2#3{{\it Zh. Eksp. Teor. Fiz. }{\bf #1} (#2) #3}
\def\jmp#1#2#3{{\it J. Math. Phys. }{\bf #1} (#2) #3}
\def\yf#1#2#3{{\it Yad. Fiz. }{\bf #1} (#2) #3}

\end{document}